\documentstyle[12pt,epsf]{article}

\def\lsim{\mathrel{\raise.2ex\hbox{$<$}\hskip-.8em\lower.9ex\hbox{$\sim$}}}
\def\gsim{\mathrel{\raise.2ex\hbox{$>$}\hskip-.8em\lower.9ex\hbox{$\sim$}}}

\renewcommand{\thesubsubsection}%
{\arabic{section}.\arabic{subsection}.\alph{subsubsection}}

\begin{document}

\font\fortssbx=cmssbx10 scaled \magstep2
\hbox to \hsize{
\includegraphics{uwlogo.ps}
\hskip.5in \raise.1in\hbox{\fortssbx University of Wisconsin - Madison}
\hfill$\vcenter{\hbox{\bf MADPH-95-915}
                \hbox{\bf astro-ph/9512079}
                \hbox{November 1995}}$ }

\vspace{.5in}

\begin{center}
{\large\bf
THE SEARCH FOR NEUTRINO SOURCES\\[3mm]
 BEYOND THE SUN}\\[1cm]
S. Barwick\\
{\it Department of Physics and Astronomy, Irvine, CA 92717}\\[2mm]
F. Halzen\\
{\it Department of Physics, University of Wisconsin, Madison, WI 53706}\\[2mm]
P.B. Price\\
{\it Department of Physics, University of California, Berkeley, CA 94720}
\end{center}

\let\Large=\normalsize
\let\large=\normalsize
\vfil

{\narrower\small
\noindent
 The hope is that in the near future neutrino astronomy, born with the
identification of
thermonuclear fusion in the sun and the particle processes controlling the fate
of a nearby supernova, will reach throughout and beyond our Galaxy and make
measurements relevant to cosmology, astrophysics, cosmic-ray and particle
physics. The construction of a high-energy neutrino telescope requires a huge
volume of  very transparent, deeply buried material such as ocean water or ice,
which acts as the medium for detecting the particles. The
AMANDA\cite{amandacoll} muon and neutrino telescope, now operating 4 strings of
photomultiplier tubes buried in deep ice at the South Pole, is scheduled to be
expanded to a 10-string array. The data collected over the first 2 years cover
the 3 basic modes in which such instruments are operated: i)~the burst mode
which monitors the sky for supernovae, ii)~ the detection of electromagnetic
showers initiated by PeV-energy cosmic electron neutrinos, and iii)~muon
trajectory reconstruction for neutrino and gamma-ray astronomy.  We speculate
on the possible architectures of kilometer-scale instruments, using early data
as a guideline.
\par}

\thispagestyle{empty}

\newpage
\setcounter{page}{1}

\section{\uppercase{High Energy Neutrino Astronomy: Science Reach}}

Attempts to push astronomy beyond the GeV photon energy of satellite-borne
telescopes, to wavelengths smaller than $10^{-16}$cm, have been initiated over
the last several decades. Doing gamma-ray astronomy at TeV energies and beyond
has turned out to be a formidable challenge. Not only are the fluxes small,
they are buried under a flux of cosmic-ray particles which is larger by
typically two orders of magnitude. Detection by air-Cherenkov telescopes of the
emission of TeV gamma rays from the Crab supernova remnant and from a pair of
nearby active galaxies has proven that the problems are not insurmountable.
Efforts are also underway to probe the sky in the corresponding energy region
by detecting neutrinos. The information from both observations should nicely
complement each other. The case for neutrino astronomy has been reinforced by
the recent realization that TeV gamma rays are efficiently absorbed on
interstellar light, rendering the Universe opaque for all but the very closest
sources. In general, high-energy photons, unlike weakly interacting neutrinos,
do not carry information on any cosmic sites shielded from our view by more
than a few hundred grams of intervening matter. Hopefully, as exemplified time
and again, the development of a novel way of looking into space invariably
results in the discovery of unanticipated phenomena.

Are there cosmic sources of high-energy neutrinos? In heaven, as on Earth,
high-energy neutrinos are produced in beam dumps which consist of a high-energy
proton accelerator and a target. Gamma rays and neutrinos are generated in
roughly equal numbers by the decay of pions produced in nuclear cascades in the
beam dump. For every $\pi^0$ producing two gamma rays, there is a $\pi^+$ and
$\pi^-$ decaying into a $\mu$ and a~$\nu_\mu$. If the kinematics is such that
muons decay in the dump, more neutrinos will be produced. We want to stress
that in efficient cosmic beam dumps with an abundant amount of target material,
high-energy photons may be absorbed before escaping the source. Laboratory
neutrino beams are an example. Therefore, the most spectacular neutrino sources
may have no counterpart in high-energy gamma rays.

\subsection{Guaranteed Cosmic Neutrino Beams from Cosmic Ray Interactions}

By their very existence, high-energy cosmic rays guarantee the existence of
sources of high-energy cosmic neutrinos\cite{PR}. Cosmic rays
represent a beam of known luminosity, with particles accelerated to
energies in excess of $10^{20}$~eV. They produce pions in interactions with the
Earth's atmosphere, the sun and moon, interstellar gas in our galaxy, and the
cosmic photon background in our Universe. These interactions are the source of
calculable fluxes of
diffuse photons and neutrinos\cite{PR}. The atmospheric neutrino beam
represents a well-understood beam dump. It can be used to study neutrino
oscillations over distances of 10 to 10$^4$~km.

The study of extremely energetic, diffuse neutrinos produced in the
interactions
of the highest energy, extra-galactic cosmic rays with the microwave background
is
of special interest. The magnitude and intensity of this cosmological neutrino
flux are determined by the maximum injection energy of the ultra-high-energy
cosmic rays and by the distribution of their sources. If the sources are
relatively near, at distances of order tens of Mpc, and the maximum injection
energy is not much greater than the highest observed cosmic-ray energy
(few${}\times 10^{20}$~eV), the generated neutrino fluxes are small. If,
however, the highest energy cosmic rays are generated by many sources at large
redshift, then a large fraction of their injection energy would be presently
contained in gamma-ray and neutrino fluxes. The effect may be amplified if the
source luminosity were increasing with redshift $z$, i.e.\ if cosmic-ray
sources were more active at large redshifts --- ``bright-phase
models"\cite{berez}.

\subsection{Active Galactic Nuclei: Almost Guaranteed}

Although observations of PeV (10$^{15}$ eV) and EeV (10$^{18}$ eV) gamma rays
are controversial, cosmic rays of such energies do exist and their origin is at
present a mystery. Cosmic rays with energies up to some 10$^{14}$~eV are
thought to be accelerated by shocks driven into the interstellar medium by
supernova explosions. The Lorentz force on a particle near the speed of light
in the galactic magnetic field (${\sim} 3 \mu$G) multiplied by the extent of a
typical supernova shock (${\sim} 50$~pc) is only ${\sim}10^{17}$~eV. Our own
Galaxy is too small, and its magnetic fields too weak, to accelerate particles
to $10^{20}$~eV. This energy should require, for instance, a 100~$\mu$G field
extending over thousands of light years. Such fields exist near the
supermassive black holes which power active galactic nuclei (AGNs). This
suggests the very exciting possibility that high-energy cosmic rays are
produced in faraway
galaxies and carry cosmological information --- on galaxy formation, for
example.

Recent observations of the emission of TeV (10$^{12}$ eV) photons from the
giant elliptical galaxy Markarian 421\cite{punch} may represent confirming
evidence. Why Mrk 421? Although Mrk 421 is the closest of these AGNs, it is one
of the weakest. The reason its TeV gamma rays are detected whereas those from
other, more distant, but more powerful, AGNs are not, must be that the TeV
gamma rays suffer absorption in
intergalactic space through the interaction with background infrared photons.
The absorption is, however, minimal for Mrk 421 with $z$ as small as 0.03. In a
study of nearby galaxies the Whipple instrument detected TeV emission from the
blazar Mrk 501 with redshift $z=0.018$, a source which escaped the scrutiny of
the
Compton GRO observatory. All this strongly suggests that many AGNs may have
significant, very high-energy components, but that only Mrk 421 and 501 are
close enough to be detected by gamma-ray telescopes.
The opportunities for neutrino astronomy are wonderfully obvious. It is likely
that neutrino telescopes will contribute to the further
study of the high-energy astrophysics pioneered by space-based gamma-ray
detectors, such as the
study of gamma-ray bursts and the high-energy emission from quasars.

Powerful AGNs at distances ${\sim}100$ Mpc and with proton luminosities
${\sim}10^{45}$~erg/s or higher are clearly compelling candidates for the
cosmic accelerators of the highest energy cosmic rays. Their
luminosity often peaks at the highest energies, and their proton flux,
propagated to Earth, can quantitatively reproduce the cosmic-ray spectrum above
10$^{18}$~eV\cite{mannbier}. Acceleration of particles is by shocks in the jets
(or, possibly, in the accretion flow onto the supermassive black
hole which powers the galaxy) which are a characteristic feature of these
radio-loud, active galaxies. Inevitably, beams of gamma rays and neutrinos from
the decay of pions appear along the jets. The pions are photoproduced by
accelerated protons interacting with optical and UV photons in the galaxy which
represent a target density of 10$^{14}$ photons per cm$^{-3}$.

A simple estimate of the AGN neutrino flux can be made by assuming that a
neutrino is produced for every accelerated proton. This balance is easy to
understand once one realizes that in astrophysical beam dumps the accelerator
and production target form a symbiotic system. Although larger target mass may
produce more neutrinos, it also decelerates the protons producing them. Equal
neutrino and proton luminosities are therefore typical for the astrophysical
beam dumps considered\cite{PR} and implies that:
\begin{equation}
4\pi \int dE (E \, dN_\nu/dE) \sim L_{CR} \sim 7.2 \times 10^{-9}\rm\ erg\
cm^{-2}\ s^{-1} \;,
\end{equation}
which simply states that the sources generate 1 neutrino for each observed
high-energy cosmic ray. Conservatively, the luminosity $L_{CR}$ has been
obtained by only integrating the highest energy component of the cosmic-ray
flux above 10$^{17}$~eV. These particles, above the ``ankle" in the spectrum,
are almost certainly extra-galactic and are observed with a $E^{-2.71}$ power
spectrum. Assuming an $E^{-2}$ neutrino spectrum, the equality of cosmic-ray
and neutrino luminosities implies:
\begin{equation}
E{dN_\nu\over dE} = {1\over4\pi} {7.5\times 10^{-10}\over E\,(\rm TeV)} \, \rm
cm^{-2}\ s^{-1}\ sr^{-1} \;.
\end{equation}

The flux of Eq.~2 is at the low end of the range of fluxes predicted in models
where acceleration is in shocks in the jet\cite{mannbier} and accretion
disc\cite{szabpro,stecker}; see Fig.~1. It is clear that our estimate is rather
conservative because the proton flux reaching Earth has not been corrected for
absorption in ambient matter in the source and in the interstellar medium. The
neutrino flux corresponds to 300 upcoming muons per year in a neutrino detector
with 10$^6$~m$^2$ effective area. Model predictions often exceed this estimate
by several orders of magnitude.

\begin{center}
\epsfxsize=3.5in\hspace*{0in}\epsffile{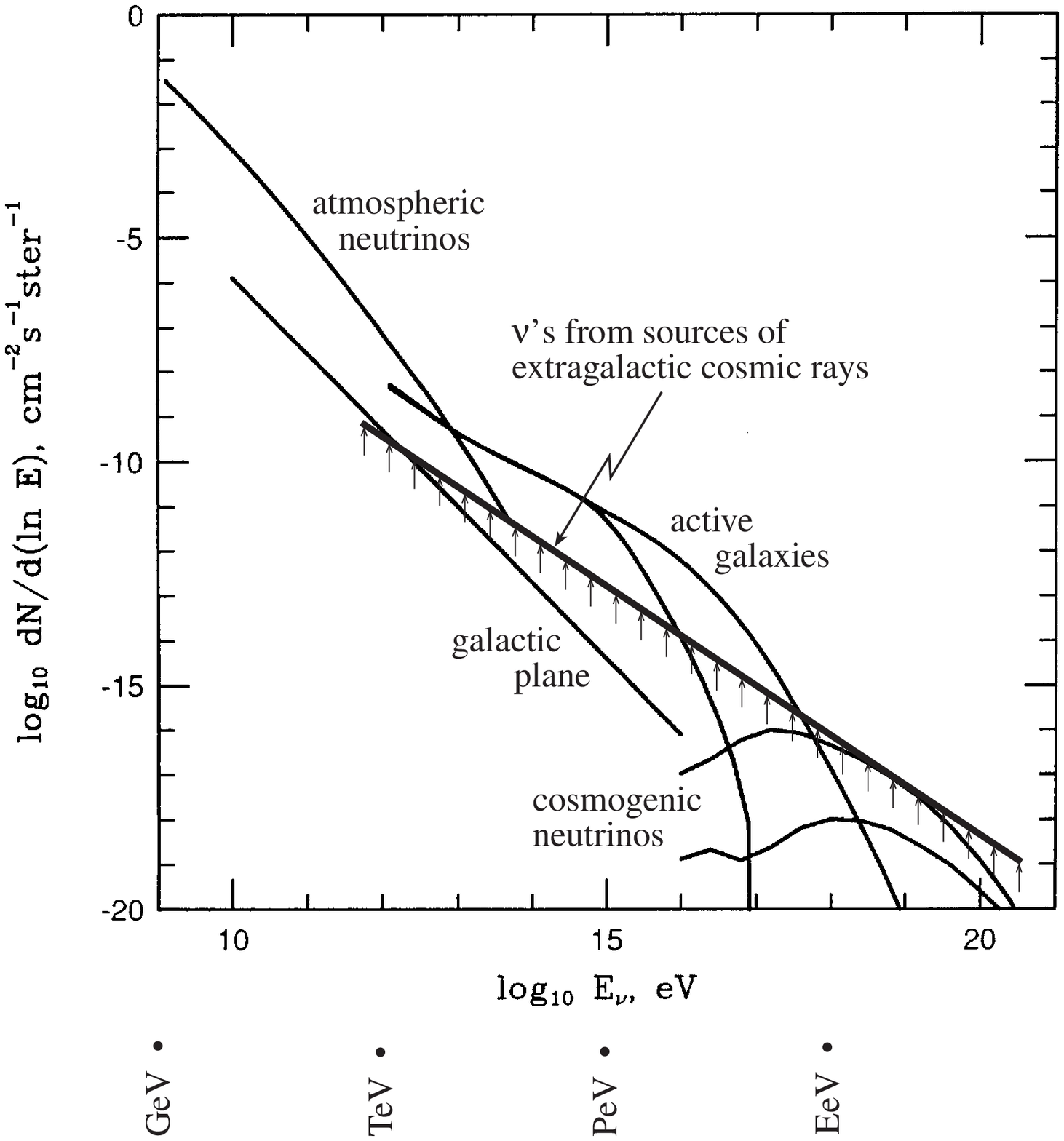}

\medskip
{\small Fig.~1: Summary of neutrino fluxes.}
\end{center}

\subsection{Interdisciplinary Aspects of High Energy Neutrino Astronomy}

The neutrino sky above 1 GeV is summarized in Fig.~1. Shown is the
flux from the galactic plane as well as a range of estimates (from generous to
conservative) for the diffuse fluxes of neutrinos from active galaxies and from
the interaction of extragalactic cosmic rays with cosmic photons. At PeV
energies and above, all sources dominate the background of atmospheric
neutrinos.

It should be emphasized that high-energy neutrino detectors are multi-purpose
instruments whose science-reach touches not only astronomy and astrophysics,
but also particle physics, cosmic-ray physics, glaciology and paleoclimatology
(oceanography) for the ice (water) telescopes. Here we enumerate issues which
represent, along with the science already discussed, high priorities in
considerations for the design and operation of high-energy neutrino detectors.

\smallskip
\noindent
{\bf i) Study of neutrino oscillations by monitoring the atmospheric neutrino
beam}\\
 Recent underground experiments have given tantalizing hints for neutrino
oscillations in the mass range $\Delta m^2 \gsim 3\times 10^{-2}\rm\
eV^2$\cite{PR}. High energy neutrino telescopes may be able to study and extend
this mass
range by measuring the zenith angle distribution of atmospheric
neutrino-induced muons. For angles of arrival of atmospheric neutrinos ranging
from vertically upward
to downward, the neutrino path length (distance from its production to its
interaction in the deep detector) ranges from the diameter of the Earth
(${\sim}10^4$~km)
to the height of the atmosphere ($\sim$10~km). With sufficient energy
resolution it is possible to observe the oscillatory behavior of the flux over
the oscillation length of several hundred kilometers suggested by the
``atmospheric neutrino anomaly".
Only a mature and well-calibrated instrument can be expected to do this
precision measurement.

\smallskip
\noindent
{\bf ii) Search for neutrinos from annihilation of dark matter particles in our
Galaxy}\\
An ever-increasing body of evidence suggests that cold dark matter particles
constitute the bulk of the matter in the Universe. Big-bang cosmology implies
that these particles have interactions of order the weak scale, i.e.\ they are
WIMPs\cite{primack}. We know everything about these particles (except whether
they really exist!). We know that their mass is of order of the weak boson
mass; we know that they interact weakly. We also know their density and average
velocity given that they constitute the dominant component of the density of
our galactic halo as measured by rotation curves. WIMPs will annihilate into
neutrinos with rates that are straightforward to estimate; {\it massive} WIMPs
will annihilate into {\it high-energy} neutrinos.

WIMP detection by high-energy neutrino telescopes is greatly facilitated by the
fact that the sun represents a dense and nearby source of cold dark matter
particles. Galactic WIMPs, scattering off protons in the sun, lose energy. They
may fall below escape velocity and be gravitationally trapped. Trapped WIMPs
eventually come to equilibrium temperature and stop near the center of the sun.
While the WIMP density builds up, their annihilation rate into lighter
particles increases until equilibrium is achieved where the annihilation rate
equals half of the capture rate. The sun has thus become a reservoir of WIMPs
which annihilate predominantly into heavy quarks and, for the heavier WIMPs,
into weak bosons. Their leptonic decays turn the sun into a source of
high-energy neutrinos with energies in the GeV to TeV range, rather than in the
keV to MeV range typical for neutrinos from thermonuclear burning. The neutrino
flux of WIMP origin is only a function of the WIMP mass. In standard cosmology
their capture and annihilation interactions are weak, and dimensional analysis
is sufficient to compute the neutrino flux from their measured density in
our galactic halo. The result is shown in Fig.~2. The interpretation of the
above arguments in the framework of supersymmetry is explicitly stated
in Ref.~\cite{PR}.

\begin{center}
\epsfxsize=3.5in\hspace*{0in}\epsffile{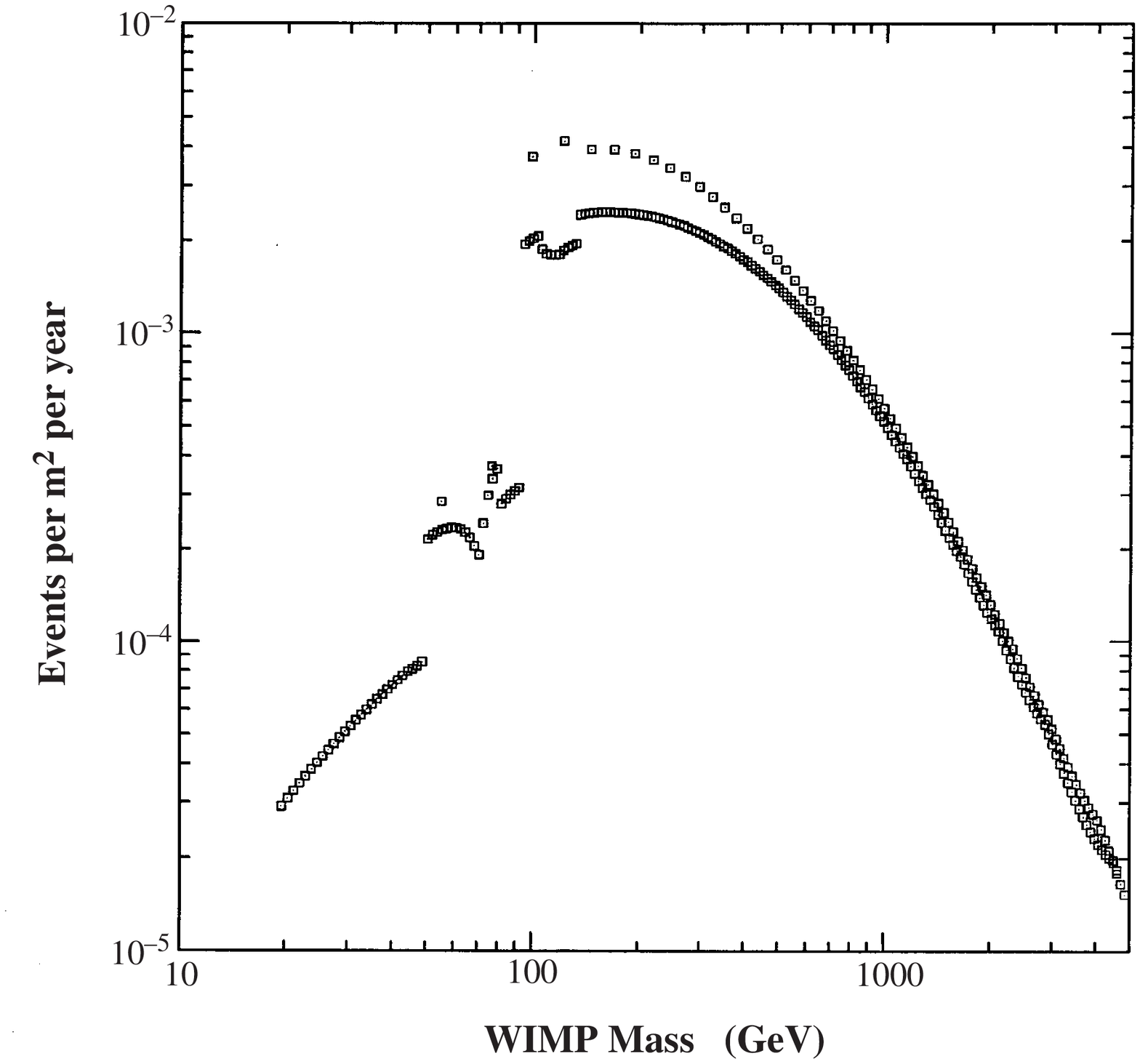}

\medskip
{\small Fig.~2: Event rates of solar, high-energy neutrinos of WIMP origin.}
\end{center}

We emphasize that experimental data, dimensional analysis and Standard
Model particle physics are sufficient to evaluate the performance of detectors
searching for such particles either directly (e.g.\ by their scattering in
germanium detectors), or indirectly (by observing their annihilation into
neutrinos in a high-energy neutrino telescope). The competing
direct method is superior only if WIMP interact coherently and their mass is
lower or comparable to the weak boson mass. In all other cases, i.e.\ for
relatively heavy WIMPs and for WIMPs interacting incoherently, the indirect
method is competitive or more powerful. For heavier WIMPS the indirect
detection
technique is especially effective and should easily extend to WIMP masses
$>$500~GeV, the upper limit reachable by future accelerators. A kilometer-size
detector probes WIMP masses well into the TeV range, beyond which they are
excluded by cosmological considerations. The rule of thumb is that a kilogram
of germanium target is roughly equivalent to a $10^4$~m$^2$ neutrino telescope.

\smallskip
\noindent
{\bf iii) Gamma-Ray Astronomy with Neutrino Telescopes}\\
The potential versatility of neutrino telescopes is dramatically illustrated by
the recent suggestion\cite{Stanev} of using neutrino detectors as gamma-ray
telescopes. Underground detectors are designed to measure the directions of
up-coming muons of neutrino origin. They can, of course, also observe
down-going muons which originate in electromagnetic showers produced by gamma
rays in the Earth's atmosphere. Although gamma-ray showers are muon-poor, it
can be shown that they produce a sufficient number of muons to detect the
sources observed by GeV and TeV telescopes. With a gamma-ray threshold higher
by one hundred and a probability of muon production by the gammas of about
$1\%$, even the shallower, lower threshold AMANDA and Lake Baikal detectors
have to overcome a $10^{-4}$ handicap. They can nevertheless match the
detection efficiency of a GeV-photon satellite detector because their effective
area is larger by a factor $10^4$.

The hundred-GeV muons observed in shallow detectors are sufficiently energetic
to leave tracks that can be adequately measured by the Cherenkov technique. The
direction of the parent photon can be inferred with degree accuracy. They
originate in TeV gamma showers whose existence has been demonstrated, at least
for two galactic and two extra-galactic sources, by air-Cherenkov telescopes. A
multi-TeV air shower will produce a 100~GeV muon with a probability of order
1\%\cite{Stanev}, sufficient to observe the brightest sources using relatively
modest size detectors with effective area of order 1000~m$^2$ or more. Although
muons from such sources compete with a large background of down-going
cosmic-ray muons, they can be identified provided the detectors achieve
sufficient effective area and angular resolution.

The key here is that for doing astronomy the muons must be sufficiently
energetic for accurate reconstruction of their direction. Very energetic muons
on the other hand are rare because they are only produced by higher energy
gamma rays whose flux is suppressed by the decreasing flux at the source and by
absorption on interstellar light. There is however a window of opportunity for
muon astronomy in the 100~GeV energy region which nicely matches the threshold
energies of the AMANDA and Lake Baikal detectors.
It is conceivable that instruments in their developing stages detect gamma-ray
sources before meeting the considerable challenge of identifying up-going muons
of neutrino origin in the large backgrounds of down-going cosmic-ray muons.

\smallskip
\noindent
{\bf iv) Supernova Search}\\
The AMANDA detector has the capability to observe the thermal neutrino emission
from supernovae\cite{SN}, even though the nominal threshold of the detector
exceeds supernova neutrino energies by several orders of magnitude. The AMANDA
4-string detector is presently monitoring our entire Galaxy and can do so over
decades in a most economical fashion. We will present details further on.

It is intriguing that, just as for the detection of AGNs, numerical
studies\cite{PR,Halzen} of the other science goals also point to the necessity
of
commissioning telescopes with at least 10$^5$~m$^2$ effective area, or more
than 10$^7$~m$^3$ volume; see e.g.\ Figs.~1,2.

\newpage
\section{\uppercase{High Energy Detectors: Area versus Threshold}}

In order to achieve the large effective detection volumes required by the
science, one optimizes the detector at high energies where: i) neutrino cross
sections are large and the muon range is increased to several kilometers, ii)
the angle between the muon and parent neutrino is less than $\sim$1 degree, and
iii) the atmospheric neutrino background is small. High energy neutrino
telescopes, just like the pioneering IMB and Kamiokande detectors, use
phototubes to detect Cherenkov light from muons, but optimize their detector
architecture to perform TeV astronomy.  Inevitably the threshold is increased
to $\sim$1~GeV from the MeV range characteristic for IMB and Kamiokande.  Such
instruments can be operated as a muon tracking device, a shower calorimeter and
a burst detector. We discuss this next.

\noindent$\bullet~$
In a Cherenkov detector the direction of the neutrino is inferred from the muon
track which is measured by mapping the associated Cherenkov cone traveling
through the detector. The arrival times and amplitudes of the Cherenkov
photons, recorded by a grid of optical detectors, are used to reconstruct the
direction of the radiating muon. For neutrino astronomy the challenge is to
record the muon direction with sufficient precision to unambiguously separate
the much
more numerous down-going cosmic-ray muons from the up-coming muons of neutrino
origin, using a minimum number of optical modules (OMs). The down-going muons
may reveal TeV gamma-ray sources, as previously discussed. Critical parameters
are the transparency of the Cherenkov medium, the depth of the detector, which
determines the level of the cosmic-ray muon background, and the noise rates in
the optical modules which will sprinkle the muon trigger with false signals.
Sources of noise include radioactive decays such as decay of potassium-40 in
water, bioluminescence and, inevitably, the dark current of the photomultiplier
tube.

\noindent
$\bullet~$
The grid of optical modules can also map PeV electromagnetic showers initiated
by
electron neutrinos, e.g.\ showers from the production of intermediate bosons in
the interactions of cosmic electron neutrinos with atomic electrons in the
detector. This technique can also detect the bremsstrahlung of very high-energy
muons of neutrino origin. Notice that {\it there is no atmospheric background
for such events once their energy exceeds 10~TeV}, although the precise value
is model-dependent; see Fig.~1. Detection of neutrinos well above this energy
would constitute the discovery of cosmic sources.

\noindent
$\bullet~$
The passage of a large flux of MeV-energy neutrinos from a supernova burst
lasting several seconds will be detected as an excess of single counting rates
in all individual optical modules of a neutrino telescope. The interaction of
$\bar\nu_e$ with hydrogen produces copious numbers of positrons with tens of
MeVs of energy.  These will yield signals in all OMs during the (typically 10
second) duration of the burst. Such a signal, even if statistically weak in a
single OM, will become significant for a sufficient number of OMs. The same
method may be used to search for gamma ray bursts provided they are, as
expected in currently favored models, copious sources of neutrinos.


\newpage
\section{BUILDING UPON FIRST DATA FROM THE\hfil\break
 FOUR-STRING AMANDA ARRAY}

\subsection{Calibration of the AMANDA Detector}

Using a hot-water drill, four strings with 20 OMs each were positioned at
depths between 800 and 1000
meters in the South Pole ice. The optical modules consist of an 8-inch EMI
9353/9351 phototube (PMT) and nothing else. The time and amplitude of the
signal are carried over a coaxial cable to the electronics
positioned at the surface above the detector. The high voltage is brought down
to the OMs on the same cable. During deployment 3 out of 80 OMs were lost. Four
other OMs, although operating, are to a varying degree problematic. Most of
these problems were associated with the first string. We adjusted our
deployment procedures and, except for a single OM, strings 3 and 4 are perfect.
Operation of the detector has been totally stable since its deployment almost
two years ago. Also deployed was a laser calibration system which pulses light
into a nylon diffuser ball positioned 30~cm below each OM. This system is fully
functional.

\looseness=-1
With 4 AMANDA strings as well as a laser calibration system in place we were
able to calibrate ice as a particle detector. Our detailed measurements
of in-situ ice\cite{inpress} exploited the laser calibration system as well as
the light emitted by cosmic-ray muons. A YAG laser was used to drive a dye
laser which pulses light of different colors into the fiber optic calibration
system. A nylon sphere deployed with each photomultiplier tube (PMT)
isotropically radiated light which was detected by other PMTs. The time
resolution of each PMT is 2~ns. By measuring the distribution of arrival times
of the pulses, both the optical properties of the ice and the position of the
tubes were accurately derived. We found that:


\smallskip\noindent
$\bullet~$
The absorption length of deep South Pole ice has the astonishingly large value
of $\sim310$~m for the 350 to 400~nm light to which the PMTs are sensitive; see
Fig.~3. A value of only 8~m had been anticipated from laboratory measurements.
For many applications the detector volume scales linearly with the absorption
length, e.g. for supernova detection. The results in Fig.~3 were first obtained
by studying the timing distribution of laser pulses at different distances of
the source. Given their importance, we have verified them with 3 independent
measurements: i) by counting photons as a function of the distance from the
laser pulse (rather than measure their timing), ii)~by studying the propagation
of the Cherenkov photons radiated by cosmic-ray muons, and iii)~by studying
coincident muon events between AMANDA and the SPASE air shower array at the
surface.

Though at first surprising, these results are understandable in terms of
conventional optics. The PMTs operate in a range of wavelengths where neither
atomic nor molecular excitations absorb the light. It seems that, in this color
interval, scattering has previously been confused
with absorption for ice as well as other transparent crystals.
Calculations of the magnitude of scattering in defect-free media and in
media containing point defects and dislocations show that the largest
contribution to scattering (with a Rayleigh $\lambda^{-4}$ dependence) is due
to dislocations that have been decorated with impurities.  The approximately
$\lambda^{-4}$ structure often seen near the minimum in published absorption
spectra of transparent solids such as LiF, NaCl, diamond, BaTiO$_3$, and ice,
is probably due to scattering from small defects, not absorption. The superb
transparency is a direct consequence of the high purity of the deep ice.

\begin{center}

\parbox[c]{2.75in}{\epsfxsize=2.75in\hspace{0in}\epsffile{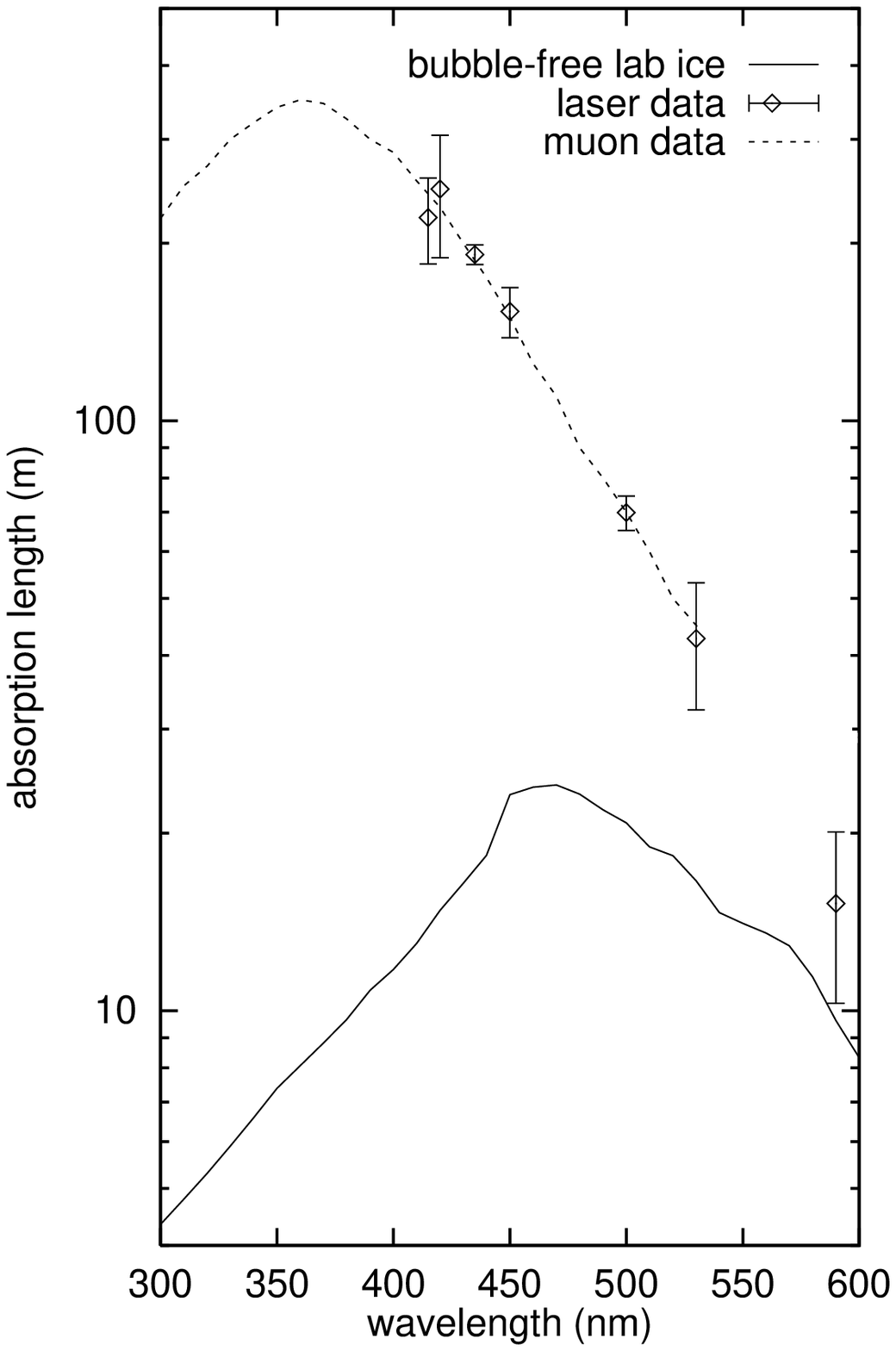}}
\qquad
\parbox[c]{3in}{\small Fig.~3: The absorption length of light in ice as a
function of  its color. Shown are the results of measurements using the AMANDA
laser calibration system (diamonds) and the Cherenkov photons from muons
(dotted line). The results for bubble-free laboratory ice are shown for
comparison.}

\end{center}

\smallskip\noindent
$\bullet~$
Ice contains residual air bubbles at 1000 m. Studies of the scattering of laser
light on residual bubbles reveal a linear decrease of their density between 800
and 1000~m in the AMANDA 4-string detector. The effective scattering length
increases from 0.25 to 1~meter at depths of 800--1000~km. At higher pressure
(greater depth) air bubbles transform indeed into a solid form of ice. Air
hydrate crystals are formed in a phase transition from hexagonal ice + air
bubbles to hexagonal ice + cubic clathrate hydrate crystals\cite{price}.
Independent of any theoretical model, microscopic studies of ice cores from
various Greenland and Antarctic sites show that bubbles and clathrate crystals
co-exist over depths of hundreds of meters but that in no case bubbles survive
to depths greater than 1550~m. Ice is bubble-free at 1250~m (Vostok), 800~m
(Dome~C), 1100~m (Byrd), 1400~m (Camp Century), and 1550~m (Dye-3). The last
two measurements are in very young Greenland ice.  With our new drilling
capabilities, future deployment beyond 1550~m should not represent a problem.

\smallskip\noindent
$\bullet~$
Ice is a sterile medium. The background noise measured in the in-situ OMs is
determined by the dark current of the photomultiplier and measured to be only
$\sim$1850 Hz, a factor 30 lower than in ocean water.


\subsection{AMANDA events: A First Glimpse at Muon Tracking, High Energy
\hfil\break
Showers and Supernova Search}

\smallskip\noindent
$\bullet~${\bf Muon trigger}.
With the complete calibration results, obtained during the 94--95 Antarctic
campaign, we have been able to simulate the performance of the detector in
detail. Reconstruction of muon trajectories in the presence of large-angle
scattering on bubbles is straightforward, now that the propagation of light in
the detector medium is adequately understood. The detector operation more
closely resembles that of a drift chamber rather than a Cherenkov detector.
Using the timing
information obtained with the laser calibration system, we determine, for each
optical module  in the trigger, the expected number of photons and their
arrival times as a function of the module's impact parameter relative to the
muon track. The expected spread in arrival time is also known. This information
is used to determine the muon direction; see Fig.~4. By simply fitting a plane
wave to events subjected to only a multiplicity cut ($>$6) and a
time-over-threshold cut, most muons can be reconstructed with sufficient
precision (better than 5~degrees in zenith angle) to obtain trigger rates and a
zenith angle distribution consistent with that expected for cosmic-ray muon
rates; see Fig.~5. One should realize here that although the timing information
is degraded by scattering, the clarity of the ice compensates as in
minimum-bias muon triggers (6 OMs on 3 strings) over 30 OMs report time and
amplitude information. More accurate reconstruction should be achieved by
refitting tracks using only OMs triggered with short times and large
amplitudes. This work is in progress.

\begin{center}
\parbox{1.5in}{\epsfxsize=1.4in\hspace{0in}\epsffile{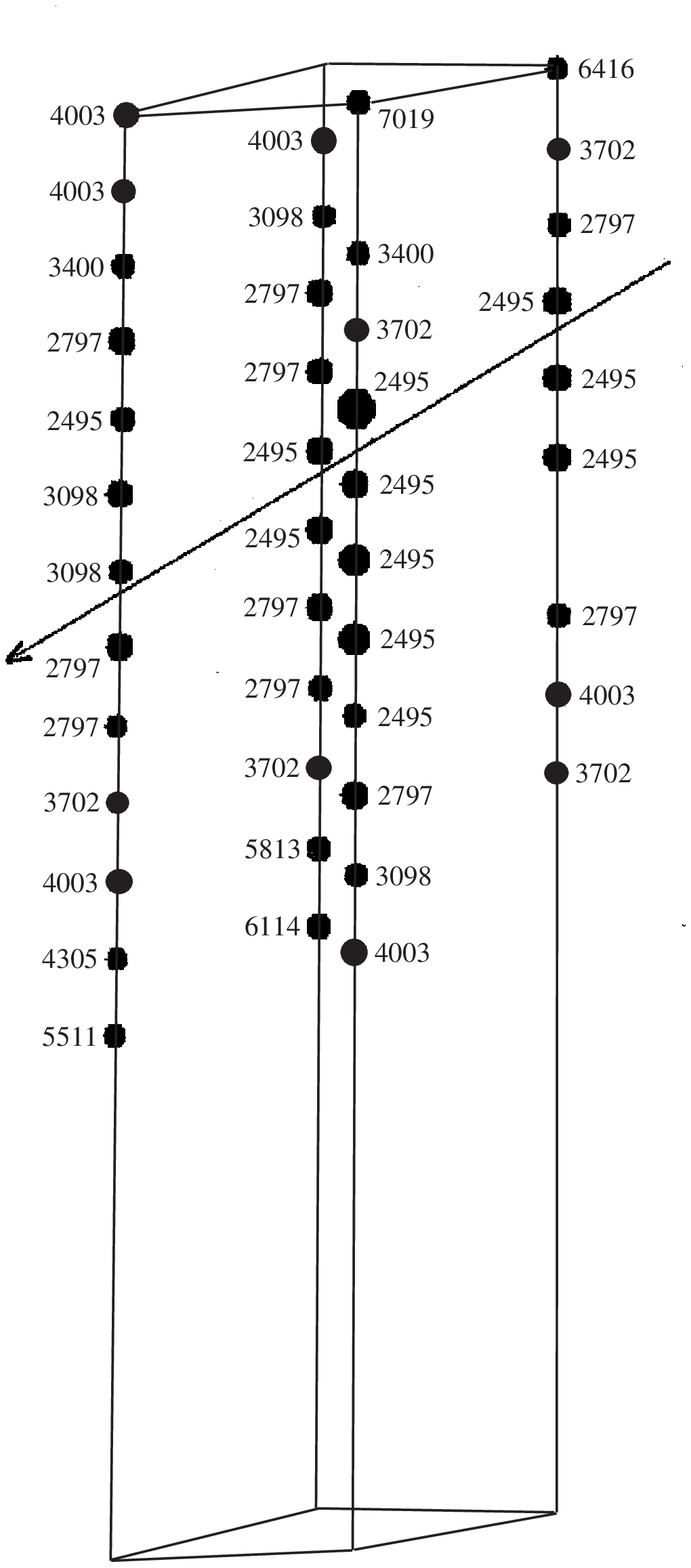}}\qquad
\parbox{3.25in}{\small Fig.~4: a) [Top left] Example of a muon bundle triggered
in  coincidence with the SPASE air shower array. The size  of the dots
represent the amplitude  for each PMT reporting in the trigger. Times are
listed next to the corresponding dot starting around 2500~ns, the time delay
between surface and deep detector.
\\
b) [Bottom left] Arrival times (in nanoseconds) and [bottom right] trigger
rates (in events per day with a given OM multiplicity) are simulated for a
range of assumptions  for the absorption length of light at the peak acceptance
of the PMTs.}
\bigskip
\epsfxsize=2.4in\hspace{0in}\epsffile{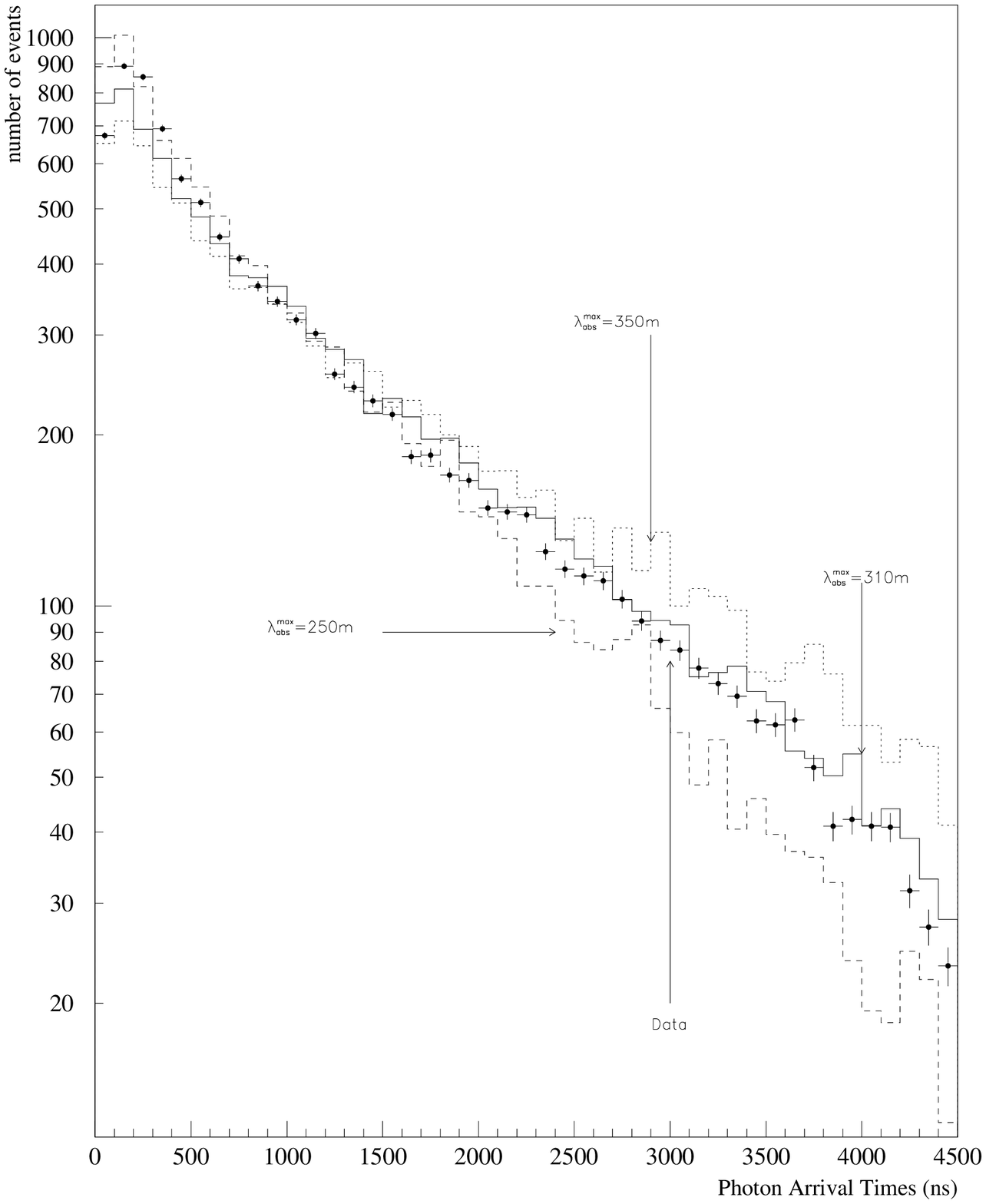}\qquad
\epsfxsize=2.8in\hspace{0in}\epsffile{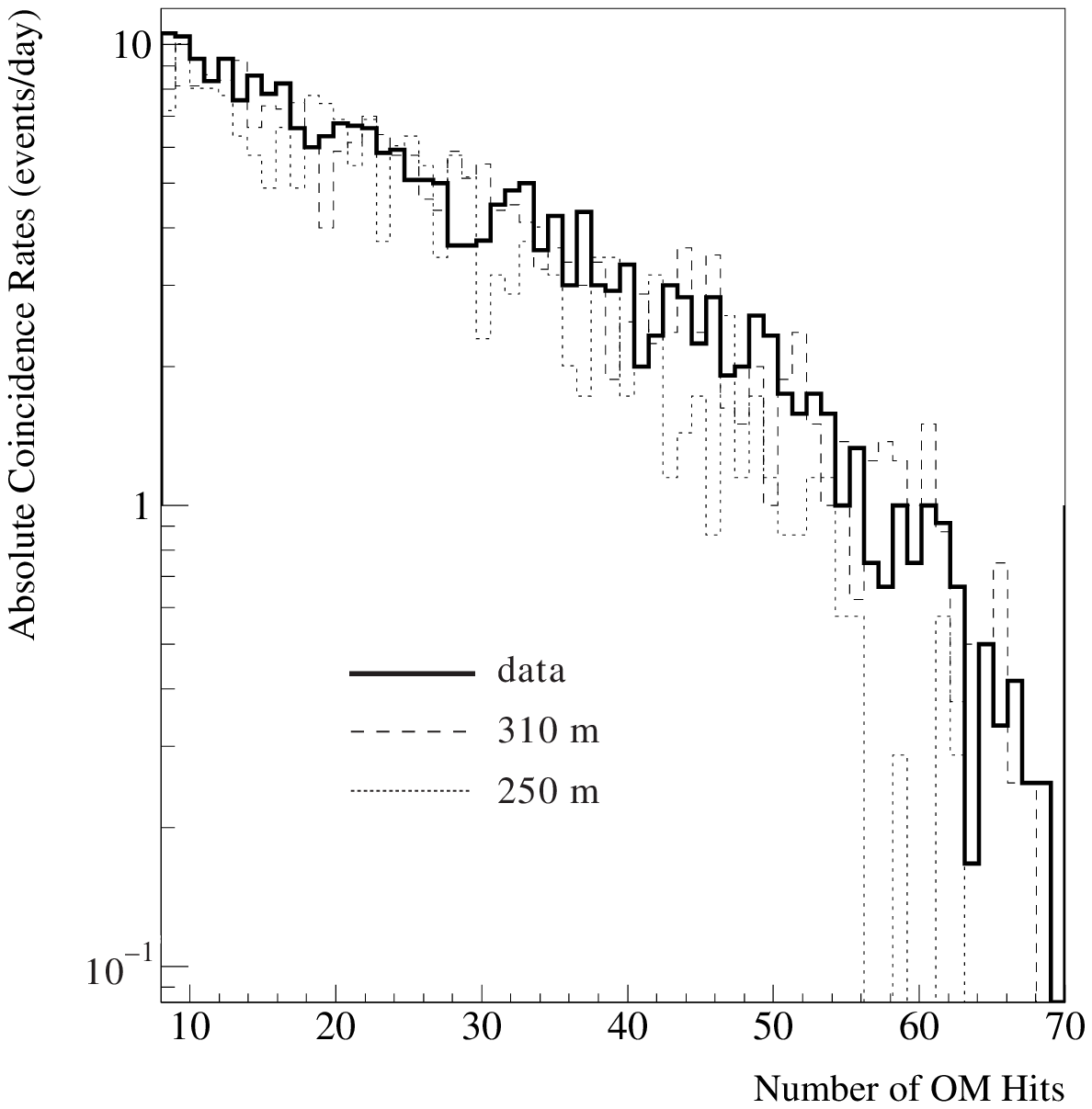}

\medskip

\end{center}

\begin{center}

\parbox[c]{2.15in}{\epsfxsize=2.15in\hspace{0in}\epsffile{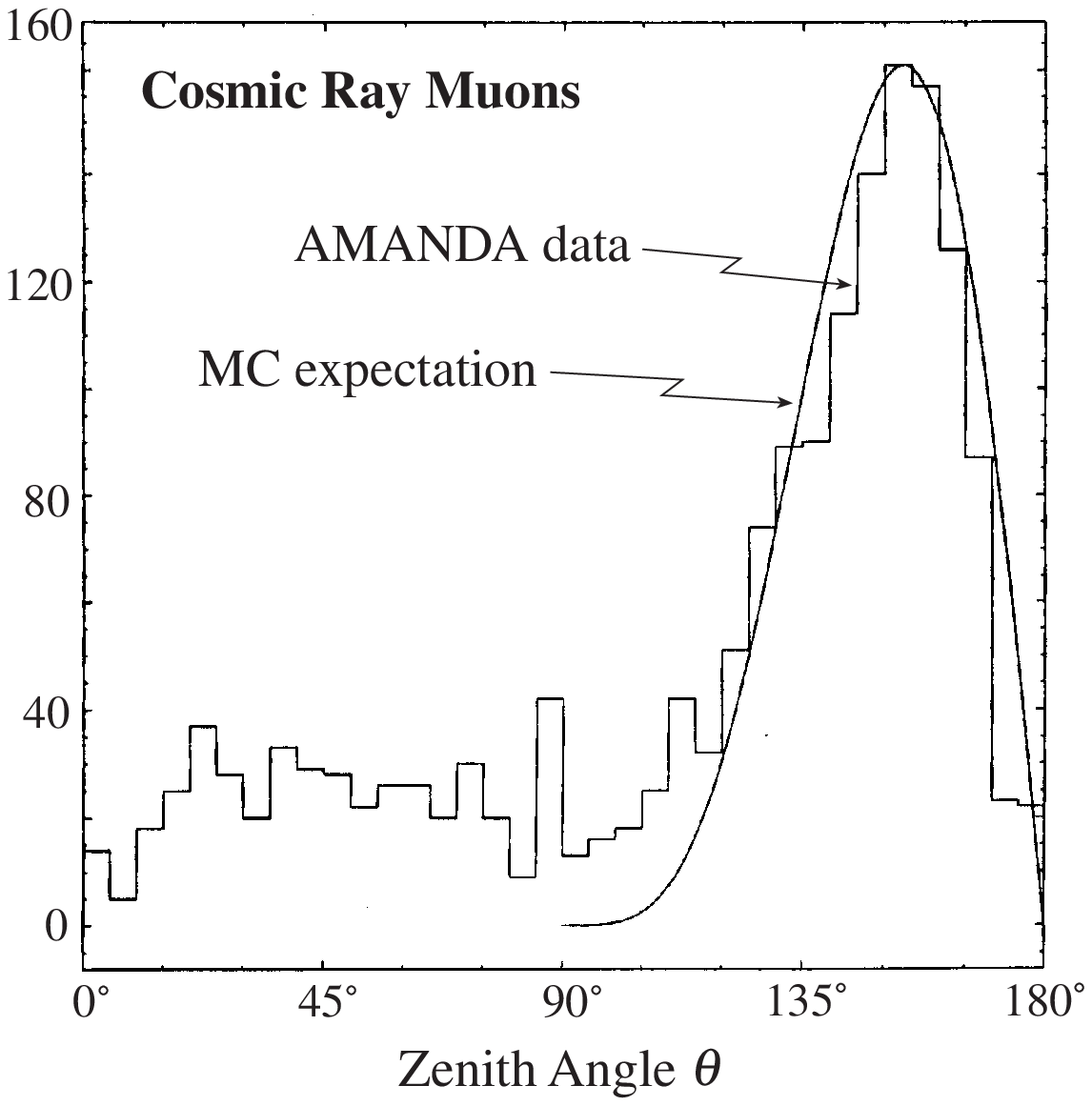}}\qquad
\parbox[c]{2.75in}{\small Fig.~5: The measured angular distribution of
cosmic-ray muons is compared to the expected one. Shown are raw AMANDA data
after a multiplicity (more  than 6 PMTs) and a time-over-threshold cut only;
track reconstruction has not  been attempted.}
\end{center}

\smallskip\noindent
$\bullet~${\bf Shower Trigger}.
The AMANDA detector represents by over an order of magnitude the largest
effective volume instrumented for the detection of PeV electromagnetic showers.
Such showers are produced by cosmic electron-neutrinos or by the radiation of
very high-energy muons. Once their energy exceeds 100~TeV the background from
atmospheric muons and neutrinos should be negligibly small for the effective
area of the present four strings; see Fig.~1. The residual bubbles cause the
Cherenkov photons from high-energy cascades to diffuse inside the detector. The
light radially propagates from the vertex of the interaction leaving a
characteristic imprint which is easy to detect and reconstruct. We have
simulated the response of the 4-string detector to 1~TeV to 10~PeV cascades by
propagating the shower photons according to simulations that quantitatively
describe the calibration measurements. We find that due to the large number of
photons generated by such cascades, the time of arrival of the first photon
detected by an OM (Leading Edge time or LE) has a small timing error provided
the cascade starts within approximately 40~meters of the PMT. The energy of an
event can be determined from a fit of LE time and of TOT (time-over-threshold)
versus distance to the PMT; see Fig.~6. The very large TOTs at intermediate
distances result from the rather broad distribution of the photon arrival time
created by the scattering on bubbles. The key here is that because of the delay
of the photons by scattering on bubbles the PMT signal now adequately
differentiates between a small signal originating nearby and a large signal far
away.

\begin{center}

\hspace{0in}\hbox{\epsfxsize=2.4in\hspace{.25in}\epsffile{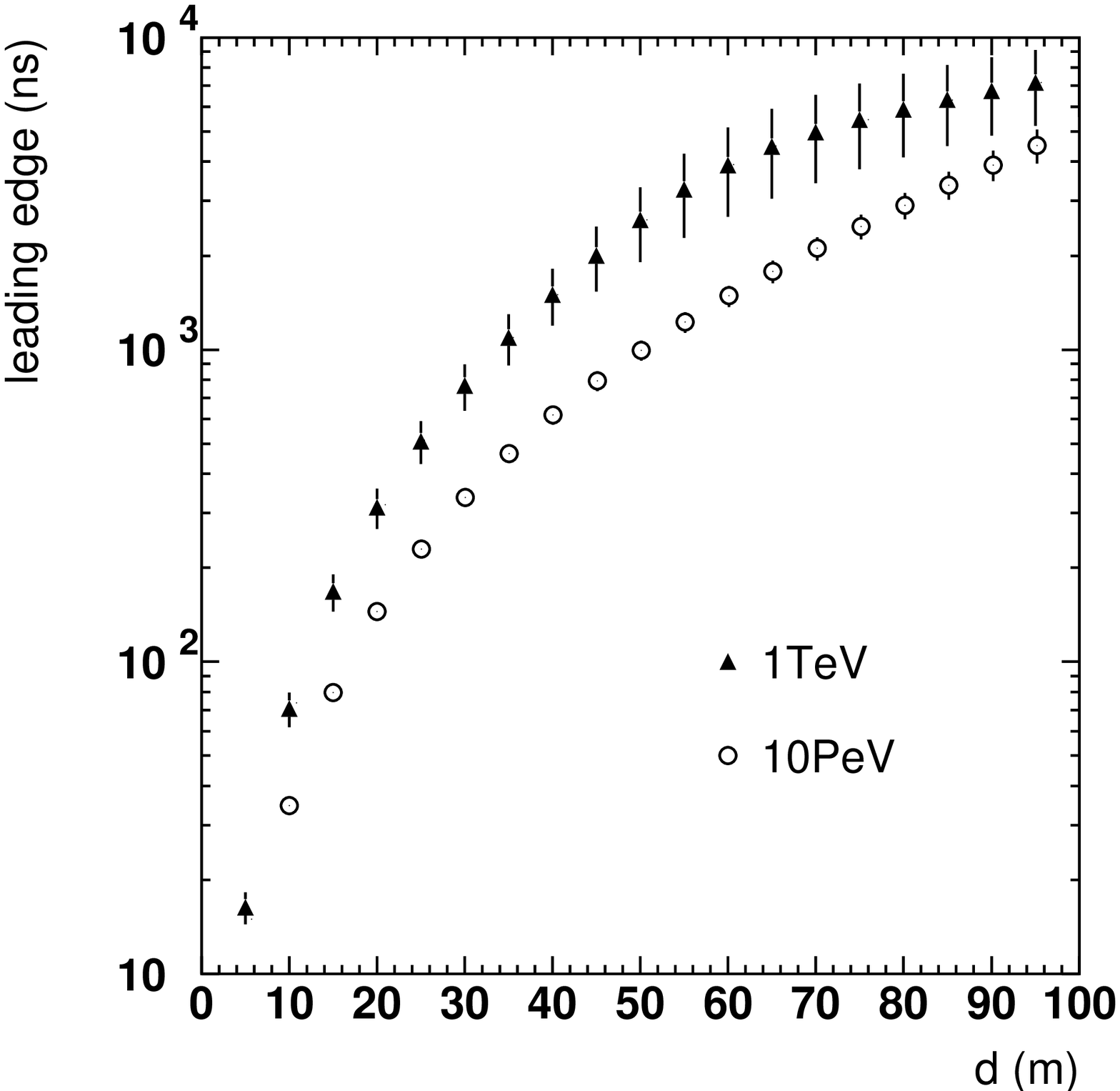}
\epsfxsize=2.4in\quad\epsffile{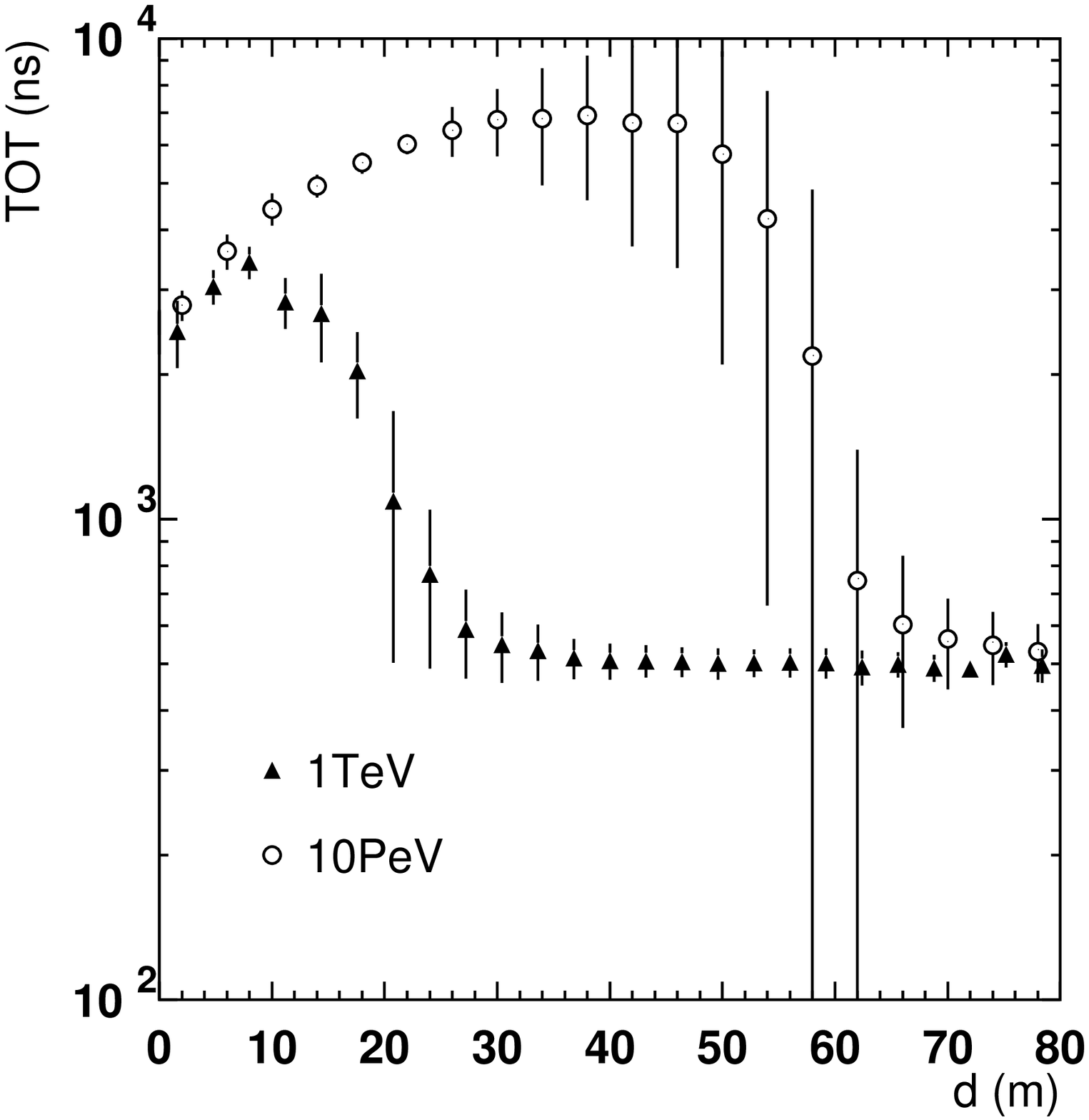}}

\medskip
\parbox{6in}{\small Fig.~6: Simulated leading edge time [left] and
time-over-threshold [right] of PMTs as a function of their radial distance to
the vertex of 1~TeV and 10~PeV cascades.}
\end{center}

Candidate cascade events are extracted from the AMANDA data stream by selecting
events with a duration of more than 5.5~microseconds that contain at least two
unusually large TOT values; a candidate event of 4~TeV energy is shown in
Fig.~7. The present detector can clearly be operated as a crude calorimeter
with an effective volume for such showers of order $10^6$~m$^3$. This may lead
to detection or improved upper limits on the fluxes of neutrinos from AGN.

\begin{center}

\mbox{
\parbox[c]{3in}{\epsfxsize=2.8in\hspace{0in}\epsffile{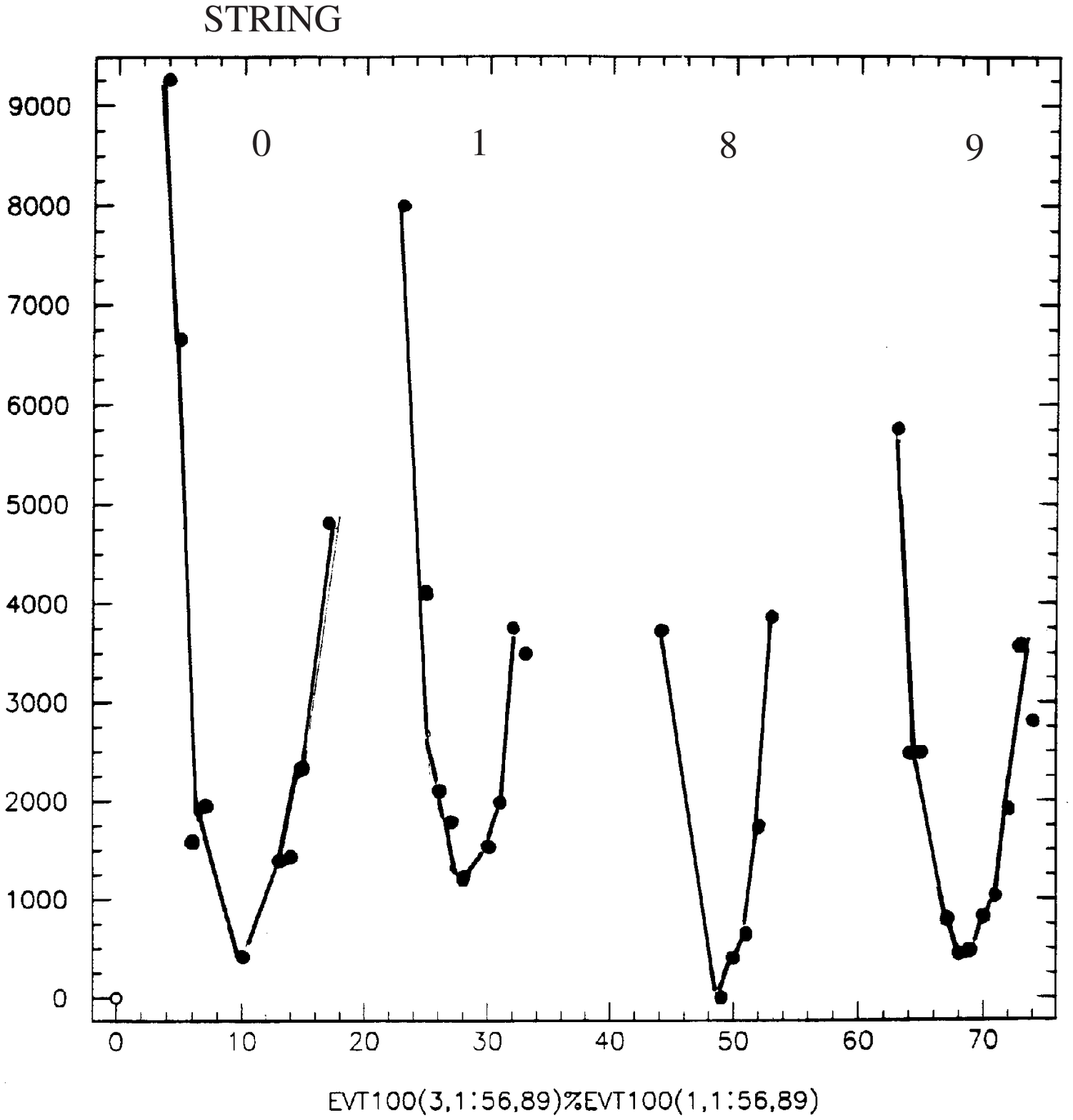}}\qquad%
\parbox[c]{3in}{\small Fig.~7: Shower event with an energy of 4~TeV. The
ordinate is PMT time (ns); the abscissa is PMT number. A shower starts at PMT
49 in string 8 and  propagates through the array. It reaches strings 0,9 at
PMTs 10,69 and, subsequently, string 1 at PMT~29. Notice the faster propagation
of light along strings in the  downward direction (larger PMT number) as a
result of the reduced density of bubbles. The event is fitted as a pure
electromagnetic shower; there is no evidence for Cherenkov emission from a muon
track which is characterized by small TOTs at short times. It is an
electron-neutrino candidate.}}
\end{center}

\smallskip\noindent
$\bullet~${\bf Supernova Trigger}.
The discovery of the large absorption length of Cherenkov light in ice has
transformed AMANDA into a supernova detector. The effective detection volume is
indeed proportional to the absorption length in the wavelength region where the
PMTs detect Cherenkov photons\cite{SN}, an increase by a factor 310/8. A
specialized trigger has been installed and detailed simulations have been
performed of the response of the detector to the stream of low-energy neutrinos
produced by a supernova. The effective radius of a module for detecting the
electrons made by supernova neutrinos is $\sim$7~m, yielding a counting rate of
300 events per optical module for the duration of a supernova at the center of
our Galaxy. This increase over the background counting rate of 1850 Hz in 73
(stable) optical modules combines to a 17~$\sigma$ observation for a galactic
supernova of the 1987A type. Theoretically, such a signal is not mimicked by a
dedicated supernova data acquisition system over the relevant time scale of
10$^2$ years. The system has been taking uninterrupted data since
February~1995. We have found that the noise distributions of the PMTs are well
described by Gaussian distribution although the width is 3 times Poissonian. A
detailed analysis of the detector's counting rate patterns is in progress.

\newpage
\section{\uppercase{What Next?}}

Even though the ultimate goal of neutrino astronomy is to commission
kilometer-size instruments, the immediate target is to demonstrate the adequate
performance of water and ice as particle detectors using a technology that can
be scaled up in a cost-effective way. For the AMANDA project the next priority
is to study the scattering length as a function of depth, especially below
1500~meters where scattering of the light is determined by the scattering on
dust, air hydrates and crystal  boundaries.
\looseness=-1
One of the advantages of building a detector in polar ice is that deployment of
OMs is not restricted to a rigid, predesigned frame. Future OMs will be
deployed below 1.5 km in order to avoid residual bubbles and improve
reconstruction of muon trajectories. (We do not exclude the possibility to
expand the kilometer-level detector as well, depending on its performance as a
shower calorimeter previously described). Larger spacings between strings, as
well as between OMs on a string, will be implemented in order to exploit the
$\sim$300~m absorption length. In Antarctic summer 95--96 six strings will be
deployed to form a pyramid in which the existing 4 strings form an apex at 0.8
to 1 km. The base, at 1.5 to 1.9 km, will consist of a large pentagon of 5
strings surrounding a central string on a circle of 40--60~meters radius; see
Fig.~8. Each string will contain 20 OMs with a vertical spacing of 20~m.

\begin{center}

\epsfxsize=2.7in\hspace{0in}\epsffile{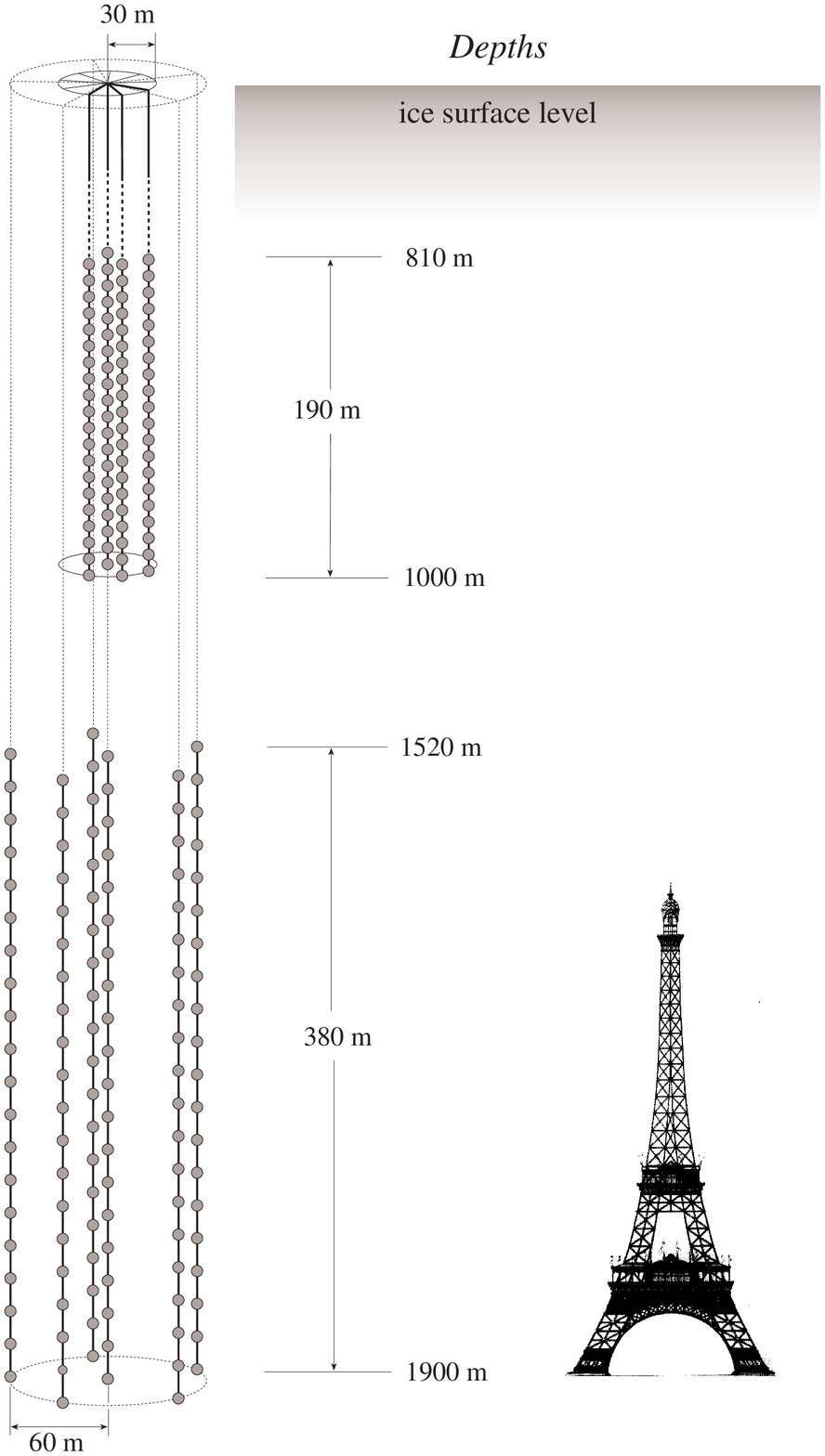}

{\small Fig.~8: AMANDA Architecture}

\end{center}

The architecture for the deployment of OMs after 1996 will, obviously, be
adjusted for any science, calibration or technological information obtained. It
is important to draw a first lesson from the initial AMANDA experience. The
present activities are dominated by muon reconstruction, mostly for attempting
gamma-ray astronomy, the search for TeV--PeV showers and the implementation of
the supernova watch. None of these topics are even mentioned in the proposal
written 4 years ago. This is exploratory science and surprises should be
expected.  The instrumentation itself, frozen into the deep ice, will not
become obsolete, and the electronics and logic located at the surface can be
updated as new ideas arise.

Future deployments will follow science as well as calibration of the ice as a
particle detector below 1~km. Given a number of OMs, design choices typically
fall between extremes: dense packing of the OMs in order to achieve good
angular resolution and low threshold, or instrumenting the largest volume of
ice in order to achieve large telescope area.


\smallskip\noindent
$\bullet~${\bf Dense-Pack Architecture}.
The first approach, pioneered by the DUMAND\cite{spiering} and
Baikal\cite{wilkes} experiments, achieves the lowest thresholds and is
therefore ideal for WIMP or neutrino oscillation searches. One can imagine
backfilling the deep detector shown in Fig.~8, especially if the scattering
length in bubble-free ice turns out to be much smaller than the absorption
length. Such detectors achieve large effective area by detecting muons far
outside the instrumented volume. The range of TeV-muons is indeed several
kilometers. A well-known handicap of these detectors is that the energy of the
muon is only determined on a logarithmic scale, e.g.\ by measuring quantities
such as the number of optical modules triggered by the muon. The resolution is
such that energy will be measured quantized in 1,10,100,1000 TeV increments.
Further problems arise because of the confusion of an energetic muon with a
bundle of low energy ones.

\smallskip\noindent
$\bullet~${\bf Distributed Architecture}.
The alternative approach where large volumes of ice are instrumented with
widely spaced OMs looks very promising, especially after the first experience
with analyzing large shower events. It emphasizes the search for the rare very
high-energy events expected from active galaxies. When the instrumented array
dimensions approach 1km, then the sensitivity to search for AGN neutrinos is
about the same whether one observes cascades or muon tracks.  For smaller
arrays this is not true since you can detect cascades only in the relatively
small instrumented volume, while you can see muons which originate from
kilometers away. This benefit is reduced
(or the challenge to reconstruct muons far outside the detector does not have
to be met) once  the array size is comparable to the characteristic muon range.
 The power to search for AGN neutrinos is now similar for muon tracks or
cascades. Also, for large arrays operated as shower detectors, scattering
represents no limit to the physics objectives as already illustrated by the
deployed AMANDA detector.

The energy resolution of such a detector may be much better that that of a
``dense-pack" instrument. Simulations of the energy resolution of the deployed
AMANDA detector using quantized 1,10,100,1000 TeV increments, suggests a
resolution of 0.25 in log($E_{measured} / E_{input}$). This assumes a Gaussian
resolution function and, at present, this has not been demonstrated. It does
suggest however that the muon energy may eventually be measured to ``a factor",
a precision unlikely to be matched by a ``dense-pack" detector.

Does this approach preclude the observation of point sources?  Probably not.
Track reconstruction is much easier once the origin of one (or more!) cascades
reveals a point (points) on the track. With only one known vertex, 2 rather
than 5 parameters are to be fitted. Track reconstruction may even work in the
presence of small scattering lengths and bubbles.

The long absorption lengths in ice really seduces one to construct a cheap
kilometer-scale detector with relatively high threshold, probably not much less
than 1~TeV. Once the instrumented volume reaches that size, the details of the
ice properties become relatively unimportant or, rather, cease to be
show-stoppers. The approach is reminiscent of the instruments detecting
acoustic or radiowave signals produced by ultra-TeV neutrinos or muons. These
methods were developed to exploit the large absorption length of acoustic and
giga-Hertz radiowaves in water or ice, allowing the deployment of detector
elements on a grid with large spacings. In the case of ice, light shares that
property. From all other points of view light has significant advantages: PMTs
represent a cheap and well-understood technology, the ambient backgrounds are
understood and the threshold of the detector is lower by one or, most likely,
several orders of magnitude.

Ice is a natural place to build ``distributed" detectors. The duration of
triggered events increases with the physical size of the detector and so does,
inevitably, the number of noise hits confusing trigger reconstruction. On a
kilometer scale this becomes a problem, especially when using large
OMs\cite{Okada}. In sterile ice the challenge is easier to meet though it is
not to be ignored.

The National Science Foundation has funded the deployment of an additional 400
OMs following the completion of the detector shown in Fig.~8. How to build up
this prototype is a complex issue, as the previous discussion illustrates. It
involves science choices. The answers may become obvious after study of the ice
properties below 1~km. If not, hybrid detectors combining both architectures
not only suggest themselves, they have already been
proposed\cite{Halzen,resvanis}

\section*{Acknowledgments}
This research was supported in part by the U.S.~Department of Energy under
Grant No.~DE-FG02-95ER40896 and in part by the University of Wisconsin Research
Committee with funds granted by the Wisconsin Alumni Research Foundation.

\end{document}